# The Characteristics of the Factors That Govern the Preferred Force in the Social Force Model of Pedestrian Movement

[1]Zarita Zainuddin, [2]Mohammed Mahmod Shuaib & [3]Ibtesam M Abu-Sulyman

***Abstract -*** The social force model which belongs to the microscopic pedestrian studies has been considered as the supremacy by many researchers and due to the main feature of reproducing the self-organized phenomena resulted from pedestrian dynamic. The Preferred Force which is a measurement of pedestrian's motivation to adapt his actual velocity to his desired velocity is an essential term on which the model was set up. This Force has gone through stages of development: first of all, Helbing and Molnar (1995) have modeled the original force for the normal situation. Second, Helbing and his co-workers (2000) have incorporated the panic situation into this force by incorporating the panic parameter to account for the panic situations. Third, Lakoba and Kaup (2005) have provided the pedestrians some kind of intelligence by incorporating aspects of the decision-making capability. In this paper, the authors analyze the most important incorporations into the model regarding the preferred force. They make comparisons between the different factors of these incorporations. Furthermore, to enhance the decision-making ability of the pedestrians, they introduce additional features such as the familiarity factor to the preferred force to let it appear more representative of what actually happens in reality.

## I. INTRODUCTION

Congestion is one of the environmental problems which have increased due to the large increase in the population growth rate. In some occasions, it has resulted in fatalities such as crowd stampede and its related problems. Solutions are urgently needed to prevent more disasters from happening. In view of this light, pedestrian studies have received much attention recently to provide solutions to these challenging problems [1]. Microscopic techniques which are basically a branch of pedestrian studies are mainly concerned with the interactions among pedestrians and their effects upon each other [1]. According to [2], the pedestrian's behavior, theoretically, can be divided into three inter-related level: 1- strategic level, where the pedestrian's activities and its order are determined; 2- tactical level, where decisions are made while performing the activities (e.g., choosing the way to an intermediate target based on the utility maximization); and 3- operational level, where instantaneous behaviors which involve most activities resulting from the interactions among pedestrians such as avoiding collision, deviation, acceleration and deceleration and other physical interactions are described. In general, researchers have considered the Social Force Model as the one which is superior among those which belong to microscopic modelling [1].

This model considers that pedestrians as self-driven particles. Apart from having the most impact and efficiency, it has also been considered as the most realistic model that can express the motivations inside pedestrians. During the last few years, researchers have conducted numerous experimental studies to compare results of this model with real life data in order to obtain more accurate values of the parameters of the model [3]-[5]. For that reason, the model has largely gone through a lot of advances. A brief demonstration of these advances has been introduced in the next section. In the third section, we have given more details about the development of the preferred force. Subsequently, we have made a comparison between the most important contributions to the preferred force. Lastly, we have incorporated a new factor into this force which is called the familiarity factor.

## II. THE SOCIAL FORCE MODEL

The Social Force Model which was originally proposed by Helbing and Moln'ar [6] is based on the concepts adopted from the social fields as described by Lewin [7]. Using mathematical approach, they modelled the behavior of pedestrians as acting forces in the Newtonian equation of motion. These forces, which are called social forces, may lead to physical reaction such as acceleration or deceleration. The model was presented to consider all the behaviors at the operational level and some of the tactical behaviors.

### A. MODELLING THE MOTIVATIONS

The system of the pedestrian's environment consists of 1- pedestrians, 2- physical environment, 3- repulsive and attractive sources (pedestrians or objects such as walls or columns), 4- intermediate targets, and 5- destination. Some of these components play an unsteady role depending on which level they belong to. (Note: for brevity, the pedestrian or the individual is referred to as "he" rather than "he or she" and "him" rather than "him or her."). Given the repulsive source $j$, it would have its effect on the motion of individual $i$ by motivating the individual $i$ to avoid the source. This psychic motivation which is exerted on $i$ by $j$ is represented as a force $\vec{f}_{ij}^{\,r}(t)$ and is termed as the social repulsive force. It is formulated in [8] by

$$\vec{f}_{ij}^{\,r}(t) := A_r e^{(R_{ij} - d_{ij}(t))/B_r} \vec{n}_{ij} \qquad (1)$$

where, $A_r$ *is* a parameter representing the interaction strength, $B_r$ is a parameter called in [4], the fall-off length parameter which represents the range of the repulsive interactions (i.e. the characteristic distance of repulsion among pedestrians). It may have different values depending on the individual's culture. $R_{ij} = r_i + r_j$ is the summation

[1,2]School of Mathematical Sciences, Universiti Sains Malaysia, 11800 USM, Pulau Pinang, e-mail :[1]zarita@cs.usm.my,[2]mh_shuaib@yahoo.com.
[3]Mathematical Science Department, College of Applied Science, Umm Al-Qura University, Saudi Arabia, email: [3]ibtesam_as@uqu.edu.sa.

of the radius of two individuals $i$ and $j$; $d_{ij}(t)$ is the distance between the centers of the two individuals at time t, $\vec{n}_{ij}$ is the normalized vector pointing from individual $j$ to individual $i$. An analogy to the repulsive force, the attractive source motivates individual $i$ to orient his direction towards the attractive source. It is formulated in [8] as

$$\vec{f}_{ij}^{a}(t) = A_{att} e^{(R_{ij} - d_{ij}(t))/B_{att}} \vec{n}_{ij} \quad (2)$$

where $A_{att}$ and $B_{att}$ are parameters which are different from the parameters of the social repulsive forces of $A_r$ and $B_r$, $\vec{n}_{ij}$ is the normalized vector pointing from pedestrian $j$ to pedestrian $i$.

A main feature of the attractive motivation is the decline in its magnitude during the response time because of the diminishing interests of individual $i$ toward $j$. An analogy with this, given the repulsive source as an object such as a wall, and given the attractive source as an object such as shops or the like, the modelling of both the repulsive and attractive motivations inside $i$ against and with these objects, respectively, has been done with slight changes of variables and reasons, as follows:

$$\vec{f}_{io}^{r}(t) := A_r e^{(r_i - d_{io})/B_r} \vec{n}_{io} \quad (3)$$

$$\vec{f}_{io}^{a}(t) := A_{att} e^{(r_i - d_{io})/B_{att}} \vec{n}_{io}. \quad (4)$$

In order to obtain a more realistic model, the individual perception is considered as a weight function as suggested in [6] that takes into account the angle $\varphi_{ij}(t)$ formed between the pedestrian direction and the vector pointing from him to the source j. Based on these accounts, the model of this function was developed in [3]:

$$W(\varphi_{ij}(t)) = \left( \lambda_i + (1 - \lambda_i) \frac{1 + \cos(\varphi_{ij}(t))}{2} \right). \quad (5)$$

An individual $i$, while he is walking, prefers to walk with a certain velocity $\vec{v}_i^0(t)$ which is different from his actual velocity $\vec{v}_i(t)$. In this case he has a motivation to adapt his actual velocity to the preferred one. A force has been included to express this motivation by the following model:

$$\vec{f}_{preferred}(t) := \gamma (\vec{v}_i^0(t) - \vec{v}_i(t)), \quad (6)$$

where $\gamma = \dfrac{m}{\tau}$, $m$ and $\tau$ represents the mass and the relaxation time respectively.

B. MODELLING THE MOTION

The total motivations mentioned above are considered as psychic tension that evokes a psychic conflict inside the individual $i$. In turn, the individual $i$ will select one of the alternative behaviors based on utility maximization [6]. The decision he made will cause physical movement to pedestrian $i$. The equations of movement are modeled mathematically in the form:

$$\frac{d\vec{x}_i(t)}{dt} = \vec{v}_i(t) \quad (7)$$

$$m_i \frac{d\vec{v}_i}{dt} = \vec{f}_i + \vec{\varepsilon}_i = \vec{f}_{preffered} + \sum_j \vec{f}_{ij} + \sum_o \vec{f}_{io} + \vec{\varepsilon}_i \quad (8)$$

$$\vec{f}_{ij}(t) = \vec{f}_{ij}^{a}(t) + \vec{f}_{ij}^{r}(t) \quad (9)$$

$$\vec{f}_{io}(t) = \vec{f}_{io}^{a}(t) + \vec{f}_{io}^{r}(t), \quad (10)$$

where $\dfrac{d\vec{x}_i(t)}{dt}$ is the temporary change of the location; $\dfrac{d\vec{v}_i}{dt}$ is the acceleration created by the forces upon individual $i$ who has mass $m_i$ and $\varepsilon_i(t)$ is the fluctuation of individual $i$.

By incorporating the panic situation into the model, a new Social Force Model was developed as shown in [3],[4]. For brevity, the model of [3], [4] is referred as the HMFV as practiced in [5]. The major feature of this incorporation is the physical interaction (contact) among pedestrians which is caused mainly by the increase of the crowd density. The interaction results in the emergence of physical forces: $\vec{f}_{pushing}$ works as a body force counteracting body compression and $\vec{f}_{friction}$ works as the sliding friction force impeding relative tangential motion [4]. The equations of these forces are modelled by:

$$\vec{f}_{friction} = \kappa \eta (R_{ij} - d_{ij}) \Delta v_{ji} \vec{t}_{ij},$$

$$\eta(x) = \begin{cases} x, & x \geq 0; \\ 0, & x < 0. \end{cases}$$
(11)

$$\vec{f}_{pushing} = k \eta (R_{ij} - d_{ij}) \vec{n}_{ij}, \quad (12)$$

where $k$ is the elasticity constant, $\kappa$ is a function of the relative tangential velocity of the two pedestrians; $\vec{n}_{ij} = (n_{ij}^1, n_{ij}^2)$ is the normalized unit vector pointing from pedestrian $j$ to pedestrian $i$; $\vec{t}_{ij} = (-n_{ij}^2, n_{ij}^1)$ is the tangential unit vector orthogonal to $\vec{n}_{ij}$ and represents the direction of $\vec{f}_{friction}$; the physical forces appear in case of contact, i.e. when $R_{ij} \geq d_{ij}$. These main contributions have resulted in a new formula of the total forces exerted upon $i$

$$\vec{f}_{ij} = \vec{f}_{socialrep} + \vec{f}_{socialatt} + \vec{f}_{pushing} + \vec{f}_{frictionr}. \quad (13)$$

Given that the object is a wall, $\vec{f}_{io}$ is obtained analogous to $\vec{f}_{ij}$:

$$\vec{f}_{io} = \vec{f}_{socialrep} + \vec{f}_{socialatt} + \vec{f}_{pushing} + \vec{f}_{friction}, \quad (14)$$

$$\vec{f}_{social,rep} = A e^{(r_i - d_{io})/B} \vec{n}_{ji} . W(\varphi_{ij}(t)), \quad (15)$$

$$\vec{f}_{social,att} = A_{att} e^{(r_i - d_{io})/B_{att}} \vec{n}_{ij} . W(\varphi_{ij}(t)), \quad (16)$$

$$\bar{f}_{pushing} = k\eta(r_i - d_{io})\vec{n}_{io} , \quad (17)$$

$$\bar{f}_{friction} = \kappa\eta(R_{ij} - d_{ij})(v_i \cdot \vec{t}_{io})\vec{t}_{io} . \quad (18)$$

## III. STAGES OF DEVELOPING THE PREFERRED FORCE

### A. THE ORIGINAL MODEL

The preferred force $\bar{f}_{preferred}(t) := \gamma(\vec{v}_i^0(t) - \vec{v}_i(t))$ is influenced by the various aspects of the preferred velocity. Here we demonstrate that the most important aspects of this velocity are dependent on the situation where the individual $i$ is surrounded by and his personal characteristics. Starting with the normal situation where there is no panic or evacuation or the like, an individual $i$ wants to reach his destination. For the case that there is no restriction on the time required for reaching the destination, the preferred velocity is expected to be the one which would give the most convenience to the individual. The determination of the preferred velocity is dependent on both the characteristics of the individual and the characteristics of the walking path and the environment. With the assumption that individual $i$ is restricted to reach his destination within a certain time, during his movement (walking), it is natural that he will be also looking for convenience, hence he is looking for a uniform movement. In the case of rectilinear path toward his destination $\vec{x}_i^0$, he would like to move (walk) to reach this uniform velocity and this represents his desired (preferred) velocity $\vec{v}_i^0(t)$:

$$\vec{v}_i^0(t) := \frac{\vec{x}_i^0 - \vec{x}_i(t)}{T-t} = \frac{s_i(t)}{T-t}\vec{e}_i^0, \quad (19)$$

$$\vec{e}_i^0(t) := \frac{\vec{x}_i^0 - \vec{x}_i(t)}{\|\vec{x}_i^0 - \vec{x}_i(t)\|}, \quad (20)$$

where $\vec{e}_i^0(t)$ is the desired direction.

In other cases where the path to the destination $\vec{x}_i^0$ have the shape of a polygon, the direction $\vec{e}_i^0(t)$ will have to be oriented towards the nearest edge (intermediate target) by which the individual $i$ intends to pass.

$$\vec{v}_i^0(t) = \frac{\|\vec{x}_i^0 - \vec{x}_i^{n-1}\| + ... + \|\vec{x}_i^j - \vec{x}_i(t)\|}{T-t} \cdot \vec{e}_i^0 \quad (21)$$

$$\vec{e}_i^0(t) = \frac{\vec{x}_i^j - \vec{x}_i(t)}{\|\vec{x}_i^j - \vec{x}_i(t)\|} \quad (22)$$

where $\vec{x}_i^j$ is the next edge among $\vec{x}_i^0, \vec{x}_i^{n-1},...,\vec{x}_i^j, \vec{x}_i(t)$ which advances $\vec{x}_i(t)$.

Naturally, the individual will be exposed to many deviations and delays, and consequently, this will affect his velocity. As a result, he will have to move with his actual velocity $\vec{v}_i(t) = d\vec{x}_i(t)/dt$ which will allow him to compensate his delay or deviation from reaching the preferred velocity. According to (19), the preferred velocity in this case will be affected according to any unsystematic change between the numerator and the denominator.

Furthermore, in certain situations, the individual $i$ will encounter circumstances which force him to take a new decision that will change the subsequent intermediate target (the next edge) since several important factors have appeared that affected his walking. The behavior of the individual to respond to these factors is a major aspect which belongs to the tactical level. The successive sections will give details about these factors and its effect on the preferred velocity.

### B. THE HELBING, MOLNAR, FARKAS AND VICSEK (HMFV) CONTRIBUTION

An important feature which has been considered as a main contribution to the preferred force $\bar{f}_{preffered} = -m\frac{\vec{v} - \vec{v}_0}{\tau}$ in the HMFV model is the incorporation of a new factor, the so-called nervousness factor (panic parameter) into the model of the preferred velocity. Thus, the preferred velocity can be expressed by a linear combination of $v_i^0(0)$, the initial preferred velocity, and $v_i^{max}$, the maximum preferred velocity. Both of which are governed by the panic parameter:

$$v_i^0(t) = [1 - p_i(t)]v_i^0(0) + p_i(t)v_i^{max} ; \quad (23)$$

$$\vec{e}_i^0(t) = Norm[(1 - p_i)\vec{e}_i + p_i\langle\vec{e}_j^0(t)\rangle_i], \quad (24)$$

where $p_i(t) = 1 - \bar{v}_i(t)/v_i^0(0)$ reflects the nervousness (panic parameter); $\bar{v}_i(t)$ is the average speed in the desired direction of motion; $\langle\vec{e}_j^0(t)\rangle_i$ is the average direction of the neighbors $j$s of $i$. A great advantage of incorporating this factor is the ability for this model to take into account the various features for different dynamics in normal and panic situations. Fig. 1 below shows how the panic parameter in the HMFV model influences the magnitude of the preferred velocity, which in turn, influences the resulting motion.

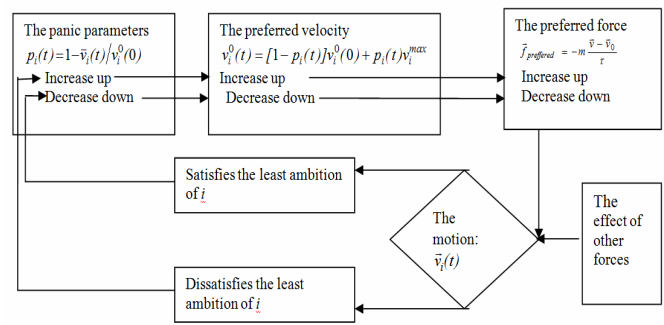

Fig.1 The above diagram shows how the panic parameter in the HMFV model influences the magnitude of the preferred velocity which in turn, influences the resulting motion.

### C. THE LAKOBA, KAUP AND FARKAS (LKF) CONTRIBUTION

The authors in [5] have claimed that the HMFV model didn't provide the individuals with any kind of intelligence or decision-making capabilities. (Note: for brevity of notations, the modified model in [5] is referred as "LKF model"). Regarding the preferred force, the modification, that has been employed in LKF model, incorporated density and memory of the locations of exits into the model. Thus, it gave the individual $i$ more independency in order to define his direction and, in turn, determine the vector of his preferred velocity:

$$\vec{e}_i = \left[ \frac{\vec{v}_i}{|\vec{v}_i|}(1-\tilde{\rho}_i) + \vec{e}_{collective}\tilde{\rho}_i \right](1-M) + \vec{n}_{i,door}M \quad (25)$$

$$\vec{e}_{collective} = \frac{\langle \vec{v}_j \rangle_i}{|\langle \vec{v}_j \rangle_i|}, \quad (26)$$

$$\vec{v}^0 = \vec{e}_i(1+E)V^0(1-D) + \langle \vec{v}_j \rangle_i D \quad (27)$$

where $M$ is the memory parameter which has the following rate of change $\frac{dM}{dt} = \frac{-M}{\tau_\pm} + \frac{\delta M(t)}{\tau_\pm}$, $\tilde{\rho}_i(t)$ indicates the non-dimensional product of the crowd density around a given pedestrian and the pedestrian area; $\vec{n}_{i,door}$ is the unit vector pointing from individual $i$ to the door; $\vec{e}_{collective}$ is the average direction of the surrounding pedestrian; $D$ is a factor that measures how the individual is dependent on others, $E$ is the individual's excitement factor which has rate of change proportional with the difference between the effective maximum excitement parameter $E_m\left(1 - \frac{v}{v^0}\right)$ and the excitement parameter itself, $E_m$ is the maximum magnitude of $E$ and lastly $V^0$ is the initial preferred force.

### IV. A COMPARISON BETWEEN THE HMFV MODEL AND THE LKF MODEL REGARDING THE PREFERRED FORCE

A comparison between the effects of each factor introduced by the last two modifications will be discussed in this section. The direction of the preferred force is the resulting direction from the addition of two vectors: the actual velocity where direction is a consequence of motion, and the preferred velocity where the direction is towards an intermediate target or a destination. Changing the direction of the preferred velocity towards another target point is it itself a tactical level behavior which is governed by the following factors. Firstly, in the HMFV model, when the individual chooses his direction he will be independent on others as long as there is no panic. In this case, because neither density nor memory factor has an apparent role, the pedestrian will keep following his direction undisturbed. In other words, if the panic parameter is low then the individualistic behavior will come into being; if it is high then the herding behavior will be the dominant behavior. However, the dependency factor $D$ has a main role in the LKF model; if the individual is completely dependent then the pedestrian would be guided absolutely by the collective direction of the others who surround him. Likewise, if he is independent, then the decision to choose a direction will be subjected to the two factors of density and memory factor. In this case the stronger the memory he has, the more stable is his direction towards the relevant exit, and the role of density is nonexistent. On the contrary, lack of memory means the density would have the main contribution to determine the direction of preferred velocity: high density will lead to a greater consideration of the collective direction of others, whereas a low one will give the individual's direction more significance.

There is almost total agreement between HMFV and LKF on the effect of the source of panic upon the magnitude of the preferred velocity, however, there are substantial differences on other issues, such as, the components of the magnitude of the preferred velocity in HMFV are weighted by the panic parameter (nervousness). Hence, an increase in the value of the panic parameter leads to amplification in the magnitude of the preferred velocity and vice versa. On the other hand, although the excitement factor, in the LKF model (which has been formed in a similar way to the formula of the panic parameter but with different modelling approach) has a similar effect on either increasing or decreasing the magnitude, it will not, however, perform in the case of a dependent individual. Thus, if the individual is dependent, then the magnitude of the preferred velocity of the individual will be the same as the collective speed of the individual's neighbors.

### V. INCORPORATING A NEW FACTOR INTO THE PREFERRED FORCE

The shortage of representing the reality with regards to modelling the preferred velocity can be deduced from the preceding discussion. Firstly, in the HMFV, the individual has no intelligence while he is in panic situation, that is, the individual has no option, other than following others, whereas, in the LKF model, the aspects of independence have been assigned to the individuals. However, this independence in LKF the model is limited by two factors: those who are independent will use their memory first to find the exit and in the case of absence of memory, they will opt to follow the majority or keep in their directions. Although in LKF model more options are available, the individuals in reality are more intelligent and have more choices to escape from a source of panic as in case of evacuation. One common aspect during evacuation is the varieties of the individuals' directions (which are more than what appear in the simulations of the last models), and these normally emerge because of the variety of the options which are available to them. The limitation of the factors of independence arises because of the simple environment of the simulations. In this section, a factor called the familiarity factor has been incorporated into the model of the preferred velocity in the LKF model to increase the options of the pedestrians to determine the direction. The function of this factor is to measure the familiarity of the pedestrian with regards to the structure of the buildings which, in turn, will influence his choices for the best route, consequently, will help him to assess which route is the safest. Hence, the direction of pedestrian $i$ is given by

$$\vec{e}_i = \left[ \left( \frac{\vec{v}_i}{|\vec{v}_i|}(1-\tilde{\rho}_i) + \vec{e}_{collective}\tilde{\rho}_i \right)(1-f) + \vec{n}_{i,route}f \right](1-M) + \vec{n}_{i,door}M$$

(28)

where $f$ denotes the familiarity factor and $\vec{n}_{i,route}$ is the the unit vector pointing from individual $i$ to the destination which is based on his assessment that it may lead to the exit; and the other denotations are the same as denoted for (25) to (27) above. The familiarity factor is assigned to each individual initially and is estimated subject to the characteristics of the environment and the different characteristics of the individual's awareness.

## VI. ACKNOWLEDGMENTS

We thank Universiti Sains Malaysia for supporting this work under the USM Fellowship.